\documentclass[twocolumn,showpacs,nofootinbib]{revtex4}
\usepackage{epsf}
\usepackage{dcolumn}

\newcommand{\BE}{\begin{equation}}
\newcommand{\EE}{\end{equation}}
\newcommand{\BA}{\begin{eqnarray}}
\newcommand{\EA}{\end{eqnarray}}
\def\be{\begin{equation}}
\def\ee{\end{equation}}
\def\bea{\begin{eqnarray}}
\def\eea{\end{eqnarray}}

\begin{document}
\input epsf
\draft
\renewcommand{\topfraction}{0.8}

\preprint{astro-ph/0307185, June 9, 2003}
\title{\bf\LARGE  Observational Bounds on Cosmic Doomsday}
\author{\bf Renata Kallosh, Jan Kratochvil, Andrei Linde}
\affiliation{ { Department
  of Physics, Stanford University, Stanford, CA 94305-4060,
USA}    }
\author{\bf Eric V. Linder}
\affiliation{Physics Division, Lawrence Berkeley National Laboratory,  Berkeley, California 94720, USA}
\author{\bf Marina Shmakova}
\affiliation{ Stanford Linear Accelerator Center, Stanford University,
Stanford CA 94309, USA} 
{\begin{abstract} Recently it was found, in a
broad class of models,  that the dark  energy density may change its sign
during the evolution of the universe. This may lead to a global collapse
of the universe within the time $t_c \sim 10^{10}-10^{11}$ years. Our
goal is to find what bounds on the future lifetime of the universe can be
placed by the next generation of  cosmological observations. As an example,
we investigate the simplest model of dark energy with a linear potential
$V(\phi) =V_0(1+\alpha \phi)$. This model can describe the present stage
of acceleration of the universe if $\alpha$ is small enough. However,
eventually the field $\phi$ rolls down,  $V(\phi)$ becomes negative, and
the universe collapses. The existing observational data indicate that the
universe described by this model will collapse not earlier than  $t_c \gtrsim 10$ 
billion years from the present moment. We show that
the data from SNAP and Planck satellites  may extend the bound on the  ``doomsday'' time to $t_c
\gtrsim 40$ billion years at the 95\% confidence level.
\end{abstract}}
\pacs{98.80.Cq, 11.25.-w, 04.65.+e,~~~~~~~~~~~~~~~ SLAC-PUB-10032,~~~~~~~~~~~~~~~~ astro-ph/0307185}

\maketitle
\section{\label{Introduction}Introduction}
Recent discovery of the acceleration of the universe
\cite{supernova,Bond,Spergel:2003cb} is often interpreted as a proof that
the universe is going to expand forever. This is indeed the case if the
acceleration occurs due to the existence of a positive vacuum energy
density $\Lambda$ (cosmological constant), constituting approximately
70\% of the energy density  of the universe $\rho_0 \sim 10^{-120} M_p^4
\sim 10^{-29}$ g/cm$^{3}$ today \cite{numbers}. However, the only existing
theoretical model based on string theory and describing an accelerating
universe in a state with a positive cosmological constant predicts that
this state is metastable, and therefore acceleration of the universe
cannot persist for an indefinitely long time  \cite{Kachru:2003aw}.

The situation becomes even more complicated in the models where the
present acceleration of the universe is related not to the cosmological
constant, but to a slowly changing energy density of a scalar field,
called dark energy. One of the first models of dark energy (and by far the simplest one)  was proposed back in 1986 in Ref. \cite{Linde1986}, where it was suggested to
replace the cosmological constant by the energy density of a slowly
changing scalar field $\phi$ with the linear effective potential
\begin{equation}\label{quint}
V(\phi) =V_0(1+\alpha \phi) \ .
\end{equation}
Here we use the units $M_p=(8\pi G)^{-1/2}=1$.  If the slope of the potential is
sufficiently small, $\alpha V_0\lesssim
10^{-120}$,  the field $\phi$ practically does not change during the last
$10^{10}$ years, its kinetic energy is very small, so at least until 
the present stage of the evolution of the universe its total potential energy
$V(\phi)$ acts nearly like a cosmological constant.
The anomalous flatness
of the effective potential in this scenario  is a standard feature of most of the models of dark energy
\cite{Banks,dark}.

The main reason to introduce this model in \cite{Linde1986} was to address
the cosmological constant problem. The main idea can be explained as
follows. Even though the energy density of the field $\phi$ in this model
practically does not change at the present time, it changed substantially
during inflation. Since $\phi$ is a massless field, it experienced
quantum jumps with the amplitude $H/2\pi$ during each time $H^{-1}$.
These jumps move the field $\phi$ in all possible directions. In the
context of the eternal inflation scenario this implies that the field
becomes randomized by quantum fluctuations: The universe becomes divided
into an infinitely large number of exponentially large parts containing all
possible values of the field $\phi$. In other words, the universe becomes
divided into an infinitely large number of `universes' with all possible
values of the effective cosmological constant $\Lambda(\phi) =V(\phi)$.
This quantity may range from $-M_p^4$ to $+M_p^4$ in different parts of
the universe, but we can live only in the `universes' with $|\Lambda|
\lesssim O(10)\rho _0 \sim 10^{-28}$ g/cm$^3$.

Indeed, if $\Lambda \lesssim -10^{-28}$ g/cm$^3$, the universe collapses
within the time much smaller than the present age of the universe $\sim
10^{10}$ years \cite{Linde:ir,Linde1986, Barrow}. On the other hand, if
$\Lambda \gg 10^{-28}$ g/cm$^3$, the universe at present would expand
exponentially fast, energy density of matter would be exponentially
small, and life as we know it would be impossible
\cite{Linde:ir,Linde1986}. This means that we can live only in those
parts of the universe where the cosmological constant does not differ too
much from its presently observed value $|\Lambda| \sim \rho _0$. This
approach proposed in \cite{Linde1986} constituted the basis for many subsequent
attempts to solve the cosmological constant problem using the anthropic
principle in inflationary cosmology
\cite{Weinberg87,Weinberg96,Garriga:1999bf,Donoghue:2000fk,Bludman:2001iz,Kallosh:2002gg,Linde:2002gj,Dimopoulos:2003iy}.

However, the simplest dark energy model with the linear potential (\ref{quint}) has a disturbing consequence:
even though the field $\phi$ moves down very slowly, eventually the
potential energy density $V(\phi)$ becomes negative, and the universe
collapses, just as the universe with a negative cosmological constant. A
detailed description of this process can be found in \cite{negative}.
Nevertheless, since the collapse will occur only in a distant future, the
linear model (\ref{quint}) with a sufficiently small $\alpha$ is quite
satisfactory from the point of view of all existing observational data.

The existence of  the vacuum instability
leading to a global collapse of the universe is a property of a large
class of the models of dark energy,
\cite{Linde1986,Banks,Kallosh:2001gr,Linde:2001ae,Kallosh:2002wj,Kallosh:2002gf,Kallosh:2003mt}.
However, it takes some time for the universe to switch from acceleration
to collapse. Recent observational data allow us to rule out many of the
models with  steep potentials predicting rapid collapse of the universe.
Most of these models  compatible with the existing observational data
predict the global collapse of the universe within the time exceeding
10-30 billion years
\cite{Kallosh:2001gr,Linde:2001ae,Kallosh:2002wj,Kallosh:2002gf,Kallosh:2002gg,Kallosh:2003mt,Alam:2003rw}.
This is the time comparable with the present age of the universe $t_0\sim
13.7$ billion years. In this respect it becomes very interesting to check
whether one could find any indication of the vacuum instability and the
future ``doomsday,'' or, {\it vice versa}, whether it is possible to
increase the ``life expectancy'' of the universe, using the most advanced
data to be obtained, for example,  by the Supernova/Acceleration Probe (SNAP: 
\cite{snap}) distance-redshift measurements of supernovae,SNAP[SN], the Planck 
Surveyor cosmic microwave background satellite \cite{planck}, 
and weak gravitational lensing from the SNAP wide field 
survey \cite{refregier}, SNAP[WL], or the LSST survey \cite{lsst}.  

 This is the main goal of our paper. We will
study this issue in the context of the simplest dark energy model
(\ref{quint}). The reason to do it is that this model (up to the field
redefinition) has only one free parameter $\alpha$, so one can relatively
easily study this model in all possible regimes. Also, this model is
quite typical: in most of the models discussed in
\cite{Kallosh:2001gr,Linde:2001ae,Kallosh:2002wj,Kallosh:2002gf,Kallosh:2002gg,Kallosh:2003mt,Alam:2003rw}
the potential looks linear with respect to $\phi$ near the point where
$V(\phi) = 0$, where the acceleration gives way to a
collapse.\footnote{A future curvature singularity may also appear in the
models where the null energy condition is violated, $\rho +p < 0$, so that
$w <-1$. We will not discuss such ``phantom'' models \cite{Caldwell:1999ew} here because their physical
interpretation is rather obscure and they are expected to lead to a very rapid
development of instability at the quantum level \cite{Carroll:2003st}.}

\section{\label{Linear Potential}Dark energy with a linear potential}

 We assume that  the  scalar field $\phi$, with the linear  potential
(\ref{quint}), represents the dark energy density of the universe. There
is also the usual matter energy density  $\rho_M= {C\over a^3}$, so that
the equations of motion are given by
 \be \ddot \phi+ 3{\dot a\over a} \dot \phi +   {\partial V\over
\partial \phi} =0 \ ,
\ee
\be
  {\ddot a\over a}= { V -\dot\phi^2-{1\over 2}\rho_M\over 3}.
\label{F2}\ee Here $a(t)$ is the scale factor of the  flat FRW metric, $
ds^2= dt^2-a(t)^2 d\vec x^2$.  The Hubble parameter is given by 
\be
H^2(t)\equiv  \left({\dot a\over a}\right)^2= {\rho_{M}(t) +
\rho_{D}(t)\over 3}\equiv {\rho_{T}(t) \over 3}\ .
 \label{F1} \ee 
We will
solve these equations numerically, more details on this can  be found in
\cite{Kallosh:2002gf}. The dark energy density $\rho_D$ and the pressure
$p_D$  are given by $ \rho_{D}= \dot\phi^2/2+ V $ and $p_{D}= \dot\phi^2/2- V
$, and the total energy density includes also the energy density of
matter, $ \rho_{T}= \rho_M+ \rho_D. $

A dimensionless dark energy (matter) density  is given by a ratio  of the
dark energy (matter) to the total energy. 
\be 
\Omega_{D}= {\rho_D\over
\rho_{T}} \ , \qquad \Omega_{M}= {\rho_M\over \rho_{T}} \ .
\ee

Observations suggest that now $\Omega_M \approx 0.28$,   $\Omega_D \approx 0.72$, and
$\Omega_T =\Omega_M  + \Omega_D   \approx 1$. The  present time will be
specified in our numerical solutions  by the moment when  $\Omega_D =
0.72$. We will study separately the effect of changing  the current value
of $\Omega_D$ between $ 0.7$ and $0.73$. Another important characteristic of the dark energy
is its pressure-to-energy ratio defining the dark energy equation of state: 
\be 
w_D= {p_D \over \rho_D}= {\dot\phi^2/2- V\over
\dot\phi^2/2+ V} \ . 
\ee 
In what follows we will drop the index $D$ in
$w_D$ and use $w$ for dark energy  $ w_D$.

Without any loss of generality one can assume that the initial value of
the field $\phi$ in the linear potential is zero, since any change of
$\phi_0$ can be absorbed into a redefinition of $V_0$.   The value of the
constant part of the potential $V_0$  for any choice of the slope is not
independent: it is chosen in a way that today at $z=0$ the value of
$\Omega_D$ equals $0.72$.  Thus there is only one independent parameter in
the linear potential model, $\alpha$, or, equivalently, the slope of the potential, $\alpha V_0$. We will assume  
that $\alpha > 0$, but all results for the lifetime of the
universe depend only on $|\alpha|$.

\begin{table*}[htbp]
\caption{ Parameters for the Models in Figs.\ \ref{scalefactor} and \ref{w1} }
\begin{tabular}{|p{60pt}|p{70pt}|p{70pt}|p{70pt}|p{70pt}|p{70pt}|}
\hline Parameter & Cosmological \par Constant & SNAP[SN]  \par + Planck 
\par + SNAP[WL]
\par ( 95\% cl ) &  SNAP[SN]  \par +Planck  \par ( 95\% cl ) &  SNAP[SN] \par
+$\sigma_{\Omega }$ \par
\par ( 95\% cl ) &
 Minimum \,Lifetime Model  \\
 \hline
 curve color& red  & orange &  purple &  blue  & black \\
 \hline
$ \alpha $& 0 &  0.71 &   0.76 &
0.86 & 1.13 \\
\hline $ V_0 $ & 0.72 $\rho_0$& 0.83 $\rho_0$& 0.85 $\rho_0$
& 0.91 $\rho_0$ & 1.77 $\rho_0$ \\ \hline
$ \alpha V_0$ & 0 &  $0.72 \times 10^{-120} M_p^3$&   $0.79 \times 10^{-120} M_p^3$&
$0.96 \times 10^{-120} M_p^3$& $2.46 \times 10^{-120} M_p^3$\\
\hline  $w(0)$   & -1 & -0.89 &  -0.87 & -0.82
&  -0.0001  \\
\hline $t_c$   &  $ \infty $ & 39.5 Gyr& 35.5 Gyr& 28.7  Gyr&
11.3 Gyr\\
\hline
\end{tabular}
\label{tab1}
\end{table*}

\begin{figure}[h!]
\centering\leavevmode\epsfysize= 5.5 cm \epsfbox{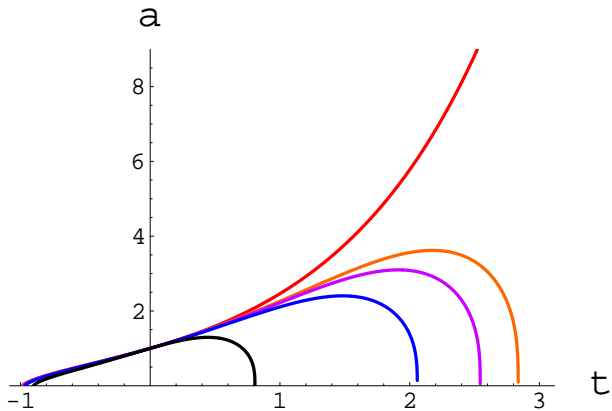}

\

\caption[fig1] {Scale factor $a(t)$ in  five models, the present moment
is $t=0$.   The upper (red) curve corresponds to the  cosmological
constant model with the vanishing slope $\alpha$; classically it has an infinite
future lifetime. The curves below (orange, purple,  blue,  and black) 
correspond to a steepening slope. The time
remaining from today to the future collapse in these models is 
shown in the table.  Time is given in units
of $H^{-1}_0 \approx 13.7/0.983$ billions of years. In these units the
current age of the universe $t \approx 13.7$ Gyr is given by $0.983$.} \label{scalefactor}
\end{figure}

\begin{figure}[h!]
\centering\leavevmode\epsfysize= 5.5 cm \epsfbox{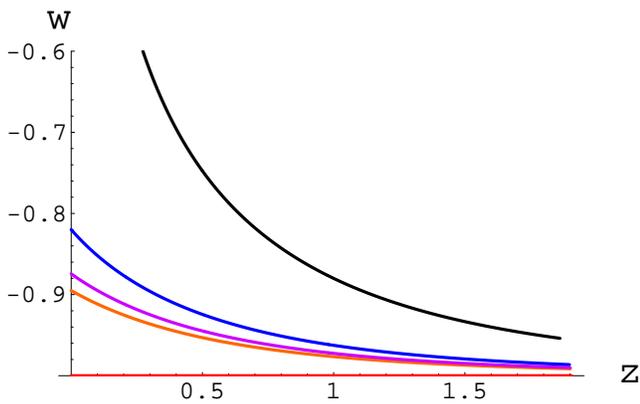}

\

\caption[w1] {Evolution of dark energy equation of state $w(z)$ in  five models;
the present moment is at $z=0$. The value $w(0)$ is also given in the table.}
\label{w1}
\end{figure}

We  solved the equation of motion of the theory numerically and we
present below a set of solutions of the linear potential model with
various slopes.  The scale factor $a(t)$ and the equations of state
$w(z)$ are plotted in Figs.\ \ref{scalefactor} and \ref{w1},
respectively.  These models are designed for the analysis of the current
and future observational data. Complete information on each model is given in the 
Table\ \ref{tab1}.
This includes the color of the curve,  the values of $\alpha$ and $V_0$, the slope $\alpha V_0$, 
the value of $w$ at present at $z=0$, and, finally, 
the time $t_c$ of the collapse of the universe  (from the present moment).

Model 1, the curve with ever expanding universe in Fig.\ \ref{scalefactor},
is a  fiducial  cosmological constant model with  vanishing slope $\alpha$;  it
has an infinite future lifetime. Models 2, 3, 4 with  increasing
slope will later be  associated with some limits on lifetime based on
specific observations. Model 5 has the largest slope $\alpha V_0=2.46 \times 10^{-120} M_p^3$, for
which the value of  $\Omega_D=0.72$ is barely reached. Any further
increase of the slope will make the model ruled out by the data  (assuming that $\Omega_D\approx 0.72$ at present) since
$\Omega_D$ in these models will never reach $0.72$.
The scale factor of the minimal model is plotted in Fig. 1, where one can
see that the universe will collapse in a time of the order $t_c=11 \mathrm{\
Gyr}$ from now.

 \begin{figure}[h!]
\centering\leavevmode\epsfysize= 5.5 cm \epsfbox{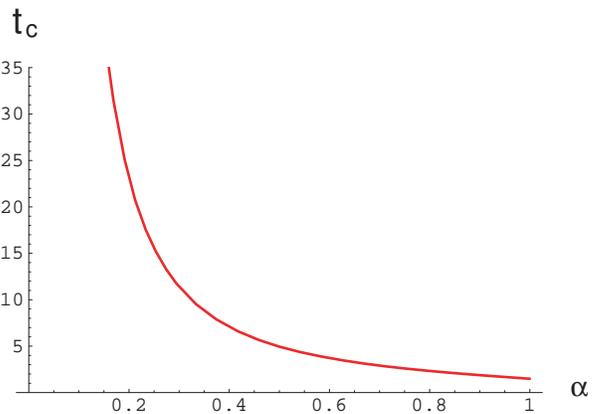}

\

\caption[fig1] {Dependence of the lifetime of the universe (starting from the present moment), in units of $H_0^{-1}$) on $\alpha$ in the linear model.  } \label{endtime}
\end{figure}

 Fig.\  \ref{endtime} shows the dependence of the time $t_c$ of the collapse of
the universe (from the present moment)  on $\alpha$.  As we
see, the lifetime sharply increases for $\alpha \ll 1$.

By looking at Figs. \ref{scalefactor}, \ref{w1} one can easily conclude
that the models with the lifetime smaller than 10 billion years are at
odds with the existing observational data. First of all, in these models
one would have $\Omega_D < 0.72$.
Secondly, 
while 
current data cannot see time variation in $w(z)$, the average value 
of $w$, defined as $\bar w = {\int da\ \Omega_D(a)\ w(a)\over  \int da\ \Omega_D(a)}$,
would be rather large, $\bar w>-0.6$.  However, for less extreme models, the 
average 
value of $w$ is generically quite close to $-1$, so data on this 
quantity 
is 
not a reliable guide.  One requires an experiment that is capable 
of 
seeing the time variation in $w(z)$.  This requires a new 
generation of 
surveys, which we discuss in the next section.

Instead of trying to find out the best constraint on the lifetime of the
universe on the basis of the present observational data, we will try to
understand to what extent the future data can allow us to predict the
fate of the universe. Our results will be limited to the simplest linear
model, but they will be quite indicative of the general situation.

\section{\label{Constraints}Constraints on the lifetime of the universe from next generation 
observations}

To constrain the dark energy models and their impact on the fate of the
universe we use observational data from     next generation cosmological
probes. The centerpiece is the Supernova/Acceleration Probe (SNAP) satellite, proposed as a dedicated dark energy mission designed to measure precise
luminosity distances to some 2000 Type Ia supernovae covering the redshift
range $z=0.1-1.7$. This data is supplemented by another part of the 
primary 
mission, 
involving wide area measurements of weak gravitational lensing.  
Valuable 
complementarity is provided by data from the Planck  Surveyor cosmic microwave
background satellite   that will use the temperature fluctuation power
spectrum to make precision determinations of several combinations of
cosmological parameters.  For our purposes the key quantity will be
measurement of the angular diameter distance to the CMB last scattering
surface at $z=1089$.
 
Of course, we do not know the results of these future experiments. They
may indicate that the universe already began decelerating, which could be
a precursor for  the future collapse of the universe. Here we would like
to study the most optimistic possibility. Let us assume that the new
observational data will favor the simplest of all possible options: the
universe is dominated by the positive cosmological constant with the
equation of state $w = -1$. In terms of our simplest model of dark energy
with a linear potential this would mean that the slope of the potential
cannot be distinguished from zero with the accuracy of the combined set
of experiments mentioned above. What kind of predictions for the future
evolution of the universe we would be able to make? In particular, we
wonder whether we will be able to say, as is often asserted, that the
acceleration of the observable part of the universe will continue
forever, and in about 150 billion years our galaxy and its closest
neighbors will remain the only inhabitants of the otherwise empty
observable part of the universe.

In order to study this question, we examine the types 
of cosmological information we will obtain.  The luminosity
and angular diameter distances are both related to the comoving distance
by factors of $1+z$.  The comoving distance is given in terms of
quantities from Section \ref{Linear Potential} by
\begin{equation}
d(z)=H_0\int_0^z {dz'\over H(z')}.
\end{equation}

 The Friedmann equations of general relativity
define the expansion history of the universe, the evolution of the scale
factor $a(t)$, in terms of the components of the energy density by
\begin{equation}
(\dot a/a)^2\equiv H^2 = H_0^2 [\Omega_M^0 (1+z)^3 + \Omega_D (z)] 
\end{equation}

The dark energy density evolves with redshift
$z=a^{-1}-1$ as
\begin{eqnarray}
\Omega_D(z) &=& \Omega_D^0 e^{3\int_0^z [dz'/(1+z')] [1+w(z')]} \\
&\to & \Omega_D^0 (1+z)^{3(1+w_0+w_a)} e^{-3w_a z/(1+z)} 
\end{eqnarray}
The second line gives the result for a commonly used parametrization
$w(z)=w_0+w_a\,(1-a)$ giving a good approximation for slowly rolling fields
\cite{linwa}.   Here $H_0$, $\Omega_M^0$,  and $\Omega_D^0$ are the
values of $H(z)$, $\Omega_M (z)$, and $\Omega_D (z)$ today at $z=0$,
respectively.

\
Since the equation of state enters via two integrals, small deviations
between the model behavior and the $w_0-w_a$ fit will be  unimportant.
This allows us to plot results in the $w_0-w_a$ space, easing
comparison
with other dark energy models.  Note that a measure of the time
variation of the dark energy equation of state is often given by
$w'\equiv dw/d\ln(1+z)|_{z=1}=w_a/2$.

Specifically, we take the SNAP baseline mission presented in \cite {klmm},
including statistical and systematic errors amounting to 1\% in distance
at the depth of the survey, $z=1.7$.  We marginalize over the absolute
magnitude parameter $\mathcal{M}$ (including the Hubble constant $H_0$)
and take a fiducial cosmological constant model with  $\Omega_D=0.72$, see
 Appendix for details.  For those cases where we
only consider supernova data, we also impose a gaussian prior on
$\Omega_D$ of 0.03; when we include CMB or weak lensing (WL) data, the natural
determination of $\Omega_D$ in complementarity with the supernovae is
better than this.  
When we incorporate WL data we use 
only information from the linear part of the mass power spectrum, in the
manner of \cite{lj03}. Note that WL is 
roughly equivalent to the CMB in complementary power with the 
supernova data.  But if the fiducial model has time varying 
equation of state $w(z)$ then there is a further gain in 
precision with all three data sets.

\begin{table*}[htbp]
\caption{Parameters for the Ellipses and Models in
 Fig.\  \ref{star72v1}}
\begin{tabular}{|p{30pt}|p{55pt}|p{55pt}|p{55pt}|p{55pt}|p{55pt}|p{55pt}|p{55pt}|}

\hline  \,  & Cosmological \par Constant & SNAP[SN] + \par Planck +\par
SNAP[WL]
\par ( 68\% cl )  &  SNAP[SN] + \par Planck +\par  SNAP[WL] \par ( 95\% cl ) &
SNAP[SN] + \par Planck \par \par ( 68\% cl )
& SNAP[SN] +\par  Planck  \par \par ( 95\% cl ) &  SNAP[SN] +
$\sigma_{\Omega }$\par
\par
\par ( 68\% cl ) & SNAP[SN] +
$\sigma_{\Omega }$\par \par
\par ( 95\% cl )   \\
 \hline
 curve color& red  & orange  dashed  & orange  &  purple  dashed &  purple
&  blue \, \, \,  dashed  &
blue  \\
 \hline
$ \alpha $& 0 &  0.576 & 0.71 &  0.63 &  0.76 & 0.72  &
0.86  \\
\hline $ V_0 $ & 0.72 $\rho_0$&  0.79 $\rho_0$& 0.83 $\rho_0$&  0.80 $\rho_0$& 0.85 $\rho_0$
& 0.84 $\rho_0$ & 0.91 $\rho_0$ \\
\hline  $w(0)$   & -1 & -0.94 &  -0.89 &  -0.92 & -0.87 & -0.89 & -0.82
  \\
  \hline $ t_c $   &  $ \infty $ & 55.3 Gyr& 39.5 Gyr& 47.9
Gyr& 35.5 Gyr& 38.6 Gyr& 28.7  Gyr\\
\hline
\end{tabular}
\label{tab12}
\end{table*}

Constraints on the dark energy model from the data
are analysed within the Fisher matrix method \cite{tegfish} which is also
explained in the Appendix. 
 The equation of state
function $w(z)$ for each model  is represented  by 2 parameters, $w_0$
and $w_a$.   The fit suggested in \cite{linwa},
 \be
\label{fit} w(z)=w_0+w_a\,(1-a)=w_0+w_a\frac{z}{1+z}  \ ,
\ee 
works well for our models,
especially for the models with larger lifetime.  Plots in the $w_0-w_a$  plane marginalize
over the value of $\Omega_D$.

\begin{figure}[h!]
\centering\leavevmode\epsfysize= 8 cm \epsfbox{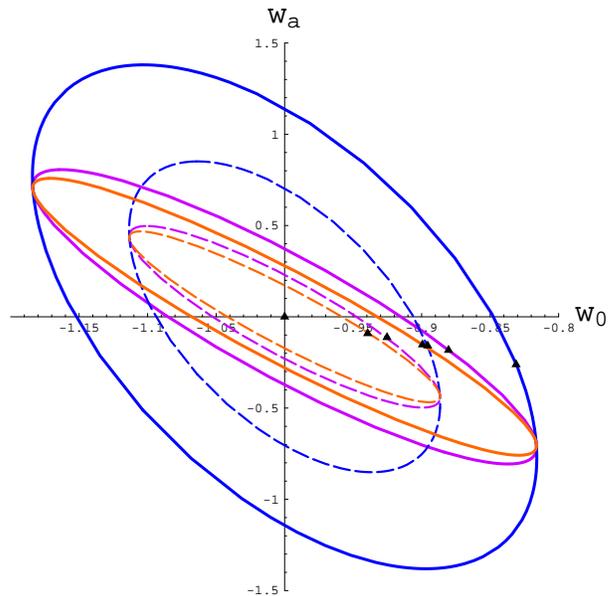} \
\caption[fig1] {Confidence contours for different combinations 
of data sets are plotted assuming a fiducial cosmological constant 
($w_0=-1$, $w_a=0$) model.  The innermost, orange (dashed) ellipse 
represents SNAP[SN] + Planck + SNAP[WL] at 95\% 
(68\%) confidence level.  The slightly wider purple ellipses use 
only SNAP[SN] + Planck, and the rounder, blue ellipses use 
only SNAP[SN].
}
\label{star72v1}
\end{figure}

Each case that we studied is shown in a $w_0-w_a$ plane as a (black) triangle in Fig.\ \ref{star72v1}. 
Each case  (apart from the cosmological constant) is chosen so
that the point in the  $w_0-w_a$ plane is  at the boundary of one of the
six ellipses, corresponding to future data. Complete information on each case
is given in  Tables I and \ref{tab12}.
The color code for the confidence ellipses, orange, purple and blue, is the same as in 
previous figures,  for SNAP+Planck+WL, SNAP+Planck, and SNAP, respectively  with 95\% confidence. 
Dashed lines in the same colors are  used for 68\% confidence data. We give the values of $\alpha$, $V_0$ and $w_0$, and
the lifetime before the collapse for all these cases.

The triangles at the boundary of each ellipse give us information about the
constraint on the lifetime before collapse in each case. Any model whose
point in the $w_0-w_a$ plane lies outside the corresponding ellipse will
be ruled out if the actual observation will favor the cosmological
constant as the most likely point in the parameter space. This will rule
out models with lifetimes before the collapse smaller than that of each
model at the boundary. When drawing the ellipses around the (future,
expected) measurement point, we made the common assumption that the
probability distribution has the same shape around the measurement value
as around the true value. This assumption is justified, as, indeed, we
confirmed that ellipses drawn around the triangles are very similar in shape
and size.

The ellipses in $w_0-w_a$ parameter space   for SNAP, for SNAP and
Planck, as well as for SNAP and
Planck and WL given in Fig.\ \ref{star72v1} are generic, model-independent, and can
be used in connection with any model---not just the linear potential
studied here---that has its $w(z)$ reasonably approximated by
(\ref{fit}). The only information that enters are the properties of the
measurements and the parametrization of the fit for $w(z)$ used. 
Therefore our results can be easily used for investigation of other models of dark energy.

The lifetimes, on the other hand, that become associated with  each point
in the $w_0-w_a$ parameter space depend on the individual model studied. The Fisher ellipses for SNAP, SNAP+Planck,  and SNAP+Planck+WL are
shown in Fig.  \ref{star72v1}.
Thus, the lifetimes that we attribute to the triangles there
are particular for the linear model.

Thus our  results can be formulated as follows.  
If the supernova distance-redshift observations from SNAP
 will yield the positive cosmological constant, i.e.\ $w_0=-1$, $w_a=0$, as
the most likely point in the $w_0-w_a$ parameter space, our analysis
allows us to rule out a future lifetime before collapse that is shorter
than $ 28.7$~Gyr at the 95\% confidence level (solid
blue ellipse), and $ 38.6$~Gyr at the 68\% confidence
level (dashed blue ellipse).

With the future Planck mission added to the SNAP data, the  corresponding
lifetime constraints can be raised even further
to $
35.5$~Gyr (solid purple) and $ 47.9$~Gyr (dashed purple)
at the 95\% and 68\% confidence levels, respectively.

Weak lensing 
similarly tightens the constraints on $w(z)$.  Both WL
and Planck successfully complement SNAP (but not each other).  The 
three data sets together lead to 
$
39.5$~Gyr (solid orange) and $ 55.3$~Gyr (dashed orange)
at the 95\% and 68\% confidence levels, respectively.

\section{ Different $\Omega_D^0$}

While all our models include $\Omega_D$ as a 
parameter to marginalize over, the fiducial, central value so 
far has been $\Omega_D^0=0.72$ as suggested  by a combination of CMB  
and large scale structure results \cite{Spergel:2003cb}
and new supernova data \cite{knop}.

\begin{figure}[h!]
\centering\leavevmode\epsfysize= 8 cm \epsfbox{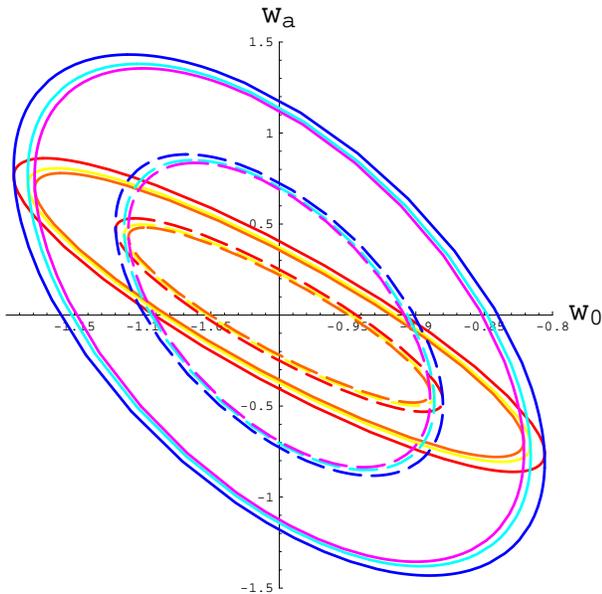} \
\caption[fig1] {Saturn plot of Fisher ellipses: Each of the Fisher ellipses shown in Fig. \ref{star72v1} is
now  given for fiducial $\Omega_D^0$ taking 3 values: $0.70, 0.72, 0.73$. The innermost
 ellipse in each case corresponds to $\Omega_D^0= 0.73$, the 
outermost one corresponds to $\Omega_D^0= 0.70$. } \label{saturn72}
\end{figure}

It is interesting to estimate 
what the difference will be in lifetimes if one changes the 
fiducial $\Omega_D^0$ to 0.7 or 0.73.
 We have constructed the Fisher ellipses  for SNAP and
SNAP+Planck for this case, see Fig.\ \ref{saturn72}. We have also evaluated
the relevant change in the lifetime bounds between   cases  $\Omega_D^0=
0.7$ and $\Omega_D^0=
0.73$. The bound is changed by few percent, therefore the changes in $\Omega_D^0$ do not lead to significant  changes in expected bounds on the lifetimes.

\section{Possible signatures of the future collapse}

Until now, we considered the possibility that the future observations will produce the simplest result, $w_0 = -1$, $w_a=0$, which would suggest that the dark energy is nothing but the cosmological constant $V_0$. In this case, because of the observational uncertainties, we would be unable to claim that the universe is going to accelerate forever, but we will be able to say, that  in the context of the simplest dark energy model (1) it will not collapse earlier than in 40 billion years from now, at the 95\% confidence level.

 \begin{figure}[h!]
\centering\leavevmode\epsfysize= 8 cm \epsfbox{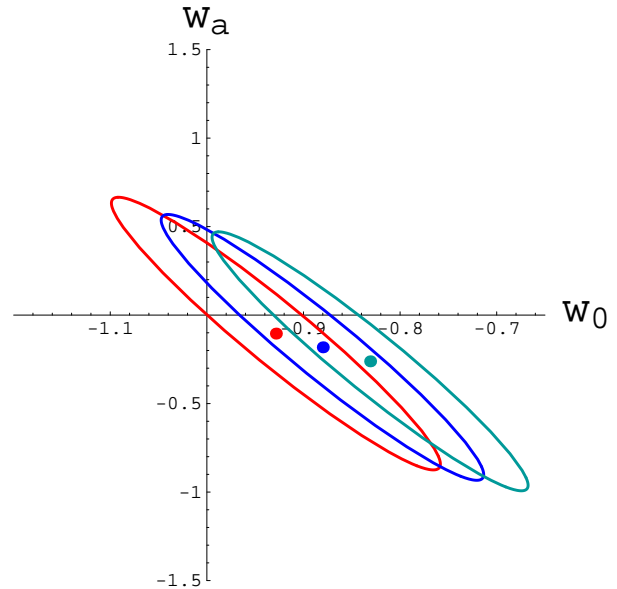} \
\caption[fig1] {Fisher ellipses with the centers corresponding to three different possible results of the future observations, including  SNAP[SN] + Planck + SNAP[WL], see the text. The points outside the ellipses are ruled out at the 95\% confidence level. The point $w_0 = -1$, $w_a=0$, corresponding to the cosmological constant, lies outside of the green and blue ellipses and at the boundary of the red ellipse.} \label{nocosmconst}
\end{figure}

But what if we get a different result? Suppose, for example, that the future observations (including SNAP[SN] + Planck + SNAP[WL]) will tell us that $w_0 = -0.83$ and $w_a = -0.26$, which corresponds to our linear model with $\alpha = 0.86$.  This would imply that at the 95\% confidence level the true values of $w_0$ and $w_a$ lie inside the green ellipse with the center at $w_0 = -0.83$, $w_a = -0.26$ (see Fig. \ref{nocosmconst}). The blue ellipse corresponds to the possibility that $w_0 = -0.88$, $w_a = -0.18$, and the red one  to $w_0 = -0.93$, $w_a = -0.1$.

Note that the point $w_0 = -1$, $w_a=0$, corresponding to the cosmological constant, lies outside  the green and blue ellipses and at the boundary of the red ellipse. 

This result has two different implications. First of all, if the future observations  find any of the sets of $w_0$ and $w_a$ discussed above, they will rule out the standard cosmological constant model at the 95\% confidence level (in the Fisher matrix approximation). 

Secondly, within the simplest model of dark energy, which is our linear model (1) \cite{Linde1986}, finding, e.g., that $w_0 = -0.93$ and $w_a = -0.1$, would imply, at the 95\% confidence level, that our universe will not exist forever, but is going to collapse. This would not be a definite proof of the coming collapse, because there may be other, more complicated models of dark energy with similar values of $w_0$ and $w_a$, which do not lead to a global collapse. Still, this would be a serious warning sign, which would stimulate further investigation of the fate of the universe.

\section{Discussion}

Our investigation leads us to the following set of conclusions.

First of all, now it becomes even more apparent that it is very difficult to
predict the future evolution of the universe. The standard textbook
illustrations showing that open and flat universes  slow down but expand forever, and a
closed universe collapses, recently became replaced by the picture of a
flat eternally accelerating universe. Now we are coming to a realization
that the stage of acceleration may be transient, and even a flat universe
may experience global collapse within a time comparable with its present
age.

Our investigation shows that even the best experiments to be carried out
in the next decade probably will be unable to give us a final answer
concerning the destiny of the universe. Even if all experiments will
unambiguously support the simplest possibility that dark energy is
nothing but a positive cosmological constant with $w(z) =-1$, this will
not really mean, as  often claimed in the popular press, that in 150
billion years our galaxy and its immediate neighborhood will remain the
sole island of matter surrounded by eternally expanding empty space. For
all we know now, and for all we are going to learn in the next ten years,
we will be unable to rule out the possibility that our part of the universe
is going to collapse in the distant future.

But one can look at it from a different perspective. We live at the very
beginning of the era of precision cosmology. It is amazing that within
the next decade, by combining several different tools such as
investigation of supernovae, CMB and  weak lensing, we will be able to learn quite a lot about the possible outcome of the universe evolution. For example, if the combination of the experiments discussed in our paper will show that the most probable parameters of the dark energy correspond to the simplest cosmological constant scenario, $w_0 = -1$, $w_a =0$, we will be able to say that in accordance with the simplest theories of dark
energy, such as our linear model,  there is no imminent danger of the
global collapse at least for the next 40 billion years.

On the other hand, if these observations will favor a different set of parameters, e.g. $w_0 > -0.93$, $w_a = -0.1$, this will rule out, at the 95\% confidence level, the simplest cosmological constant scenario. This result would be of great importance for our understanding of the most fundamental issues of physics, such as the structure of the vacuum state. In addition, such a result would mean, at least in the context of the simplest model of dark energy (\ref{quint}), that our universe is going to collapse. This would make further investigation of dark energy even more urgent and interesting.

\

It is a pleasure to thank S. Church,
S. Dimopoulos, M. Peskin, A. Starobinsky, and S. Thomas for useful discussions 
and Pisin Chen for support of this work. The
work by  R.K., J.K. and A.L. was supported by NSF grant PHY-9870115.  The work by J.K. and A.L.  was also supported by the Templeton
Foundation grant No. 938-COS273. The work by M.S. was supported by DOE grant  DE-AC03-76SF00515. The work by E.L. was supported in part
by the Director, Office of Science, DOE under DE-AC03-76SF00098 at LBL.

\vskip 2 cm

\section*{Appendix: SNAP, Planck, and Fisher ellipses}
This appendix is designed to offer a simple, practical
guide on the implementation of the Fisher matrix method
for analysis of data constraints on cosmological parameters.
For the mathematical basis of the method see \cite{tth}
and for a general application to cosmology see \cite{tegfish}.
Here we give a step by step introduction to allow those
unfamiliar with the method to use it immediately (also see
the appendix of \cite{klmm}).

Formally, the Fisher matrix is defined by the expectation value 
\BA
\label{Fisher-Def} F_{ij}&\equiv&\left\langle-\frac{\partial^2\ln
L(\mathbf{x},\mathbf{\bar p})}{\partial p_i\partial
p_j}\right\rangle=\nonumber\\ &=&\left\langle\frac{\partial\ln
L(\mathbf{x},\mathbf{\bar p})}{\partial p_i}\frac{\partial\ln
L(\mathbf{x},\mathbf{\bar p})}{\partial p_j}\right\rangle, 
\EA 
where
$L(\mathbf{x},\mathbf{p})=\Pi_{k=1}^Kf(x_k,\mathbf{p})$ is  the combined
probability distribution, and $f(x_k,\mathbf{p})$ is the probability
distribution of the individual measurement $x_k$ that in general also
depends on all the model parameters $\mathbf{p}=(p_1,\dots,p_N)$.
$\mathbf{\bar p}$ denotes the fiducial or (unknown) true parameter value of
$\mathbf{p}$.  The second equality in (\ref{Fisher-Def}) follows from the matrix
properties in the maximum likelihood approach, see \cite{tth} and
references therein for a derivation. In many cases the probability
distribution
$ L $  could be approximated by a Gaussian  
 and its relation to observed quantities,  
 such as supernova magnitudes,   becomes very simple \cite{tth}.

Practically, the Fisher matrix method provides a local approximation
to the likelihood surface for model parameters $p_i$
given a set of observations $\mathbf{x}=\{m_k\}$.  It involves the
sensitivities, or first derivatives, $\partial m_k/\partial
p_i$ evaluated at the fiducial model $\{\bar p_i\}$ 
\cite{tegfish}.  This
means that it only gives good estimates for small fit
uncertainties, i.e.\ where the data constraints are strong
enough to limit consideration to models near the input.
Biases as well as uncertainties can be treated within the
Fisher method.

The sensitivities combine into the symmetric Fisher matrix:
\begin{equation} \label{FisherMatrix}
F_{ij}=\sum_k \frac{1}{\sigma^2(m_k)}\frac{\partial m_k}
{\partial p_i} \frac{\partial m_k}{\partial p_j}.
\end{equation}
Here we have assumed that the covariance of the data points vanishes, so
the covariance matrix of the errors reduces to diagonal entries
$1/\sigma^2(m_k)$.  This is a reasonable approximation under certain
circumstances, such as when the data is grouped into redshift bins wider
than the correlation length.  Remember that the Fisher matrix is a quick
and rough approach; if one wants a more rigorous treatment then one might
as well use a full Monte Carlo of the data set (including covariances).

One can see that the larger the derivatives (i.e.\ more sensitivity), the
larger the matrix entries. Similarly, the more (independent) data points
or the smaller the data errors, the larger the entries.  This leads to
the Fisher matrix also being known as the information matrix and this
name gives a good guide to the interpretation.  The larger a matrix entry,
the more information the data has provided on those model parameters:
hence it is a more sensitive probe and a better final parameter estimator.

The parameter estimations and their covariances can be read off from the
inverse of the Fisher matrix, known as the error or covariance matrix
$C=F^{-1}$.  So for example $C_{ii}=\sigma^2_{ii}$ gives the one sigma
error (68\% confidence level) on the parameter $p_i$, averaged over all
the other dimensions of the parameter space. Formally, the Fisher method
only gives a lower limit on this error, as given by the Cram\'er-Rao
inequality \BE \label{Cramer-Rao} \Delta p_i\geq\frac{1}{\sqrt{F_{ii}}},
\EE where $\Delta p_i\equiv\sqrt{\langle p_i^2\rangle-\langle
p_i\rangle^2}$  is the variance of the model parameter $p_i$, given 
the measurement errors. The Cram\'er-Rao lower bound theorem
(\ref{Cramer-Rao}) is proved readily from the general property \BE
\mathrm{Var}(X)\geq\frac{(\mathrm{Cov}(X,Y))^2}{\mathrm{Var}(Y)}, \EE
which holds for any two random variables $X$ and $Y$, by taking
$X\equiv p_i(x_1,...,x_K)$, $Y\equiv\frac{\partial\ln(L(x_1,\dots,x_K;
p_1,\dots,p_N))}{\partial p_i}$, and using that, for these $X$ and $Y$,
we have $\langle Y\rangle=0$, $\mathrm{Cov}(X,Y)=1$ and \BE
\mathrm{Var}(Y)=\left\langle-\frac{\partial^2\ln(L(x_1,\dots,x_K;p_1,\dots,p_N))}{\partial
p_i^2}\right\rangle. \EE

One can deal with the full set of parameters in two basic ways: by fixing
those not of immediate interest or by averaging over their probability
distribution.  The first is accomplished practically by excising the rows
and columns corresponding to the fixed parameters from the Fisher matrix
before inverting it to find the error matrix.  This is a severe step,
equivalent to assuming perfect knowledge of that variable. The second
approach, known as marginalization, excises the appropriate rows and
columns from the covariance matrix.  Thus, as stated before, $C_{ii}$
gives the error on a single parameter, averaging over all others.

If one has multiple data sets, then the information simply adds (if the
data sets are independent).  That is, one simply adds the Fisher matrices,
assuming they have a common parameter set (note, the $m$'s in
(\ref{FisherMatrix}) can be completely different quantities for the two
Fisher matrices added, what is important is that the $p_i$'s have exactly
the same meaning). So if one wants to constrain cosmological parameters
using both supernova distance data and CMB power spectrum data, then one
just adds the Fisher matrices of each experiment.  As a shortcut,
sometimes one directly places a parameter constraint on the phase space
rather than incorporating the data from the beginning.  This is known as
a prior, and is also implemented by adding a Fisher matrix; if the prior
is on a single parameter, e.g.\ $\Omega_D$ is known to
$\sigma_{\Omega_D}$, then the addition is of a matrix empty except for
one  entry $F_{ii}^{\mathrm{prior}}=1/\sigma^2_{\Omega_D}$, where
$i=\Omega_D$.  Strictly speaking, this should be done only if the prior
knowledge truly has no covariance with any other parameters.  If one is
interested in the effects of different priors on the final parameter
estimation, then one can use rules of matrix algebra to quickly determine
how this extra bit of information affects the results (see
\cite{astier}).  Biases, where the data has been skewed from the true
model behavior by systematic errors, i.e.\ $m_k\to m_k+\delta m_k$, can
also be treated by matrix manipulation (see \cite{hutwl} and the Appendix 
of \cite{klmm}). 

\subsection*{SNAP}
The SNAP data set will comprise high precision and accuracy measurements of
the astronomical magnitudes of some 2000 Type Ia supernovae from redshift
$z=0$ out to redshift $z=1.7$.  Magnitudes are logarithmic distance variables
and depend on the cosmological variables, say $\Omega_D$, $w_0$, and
$w_a$, and an unimportant (for cosmology) absolute supernova luminosity
variable written as ${\cal M}$.  We will always marginalize over ${\cal
M}$, assuming a uniform probability distribution.

For the level of statistical analysis considered here, the supernova
magnitude data is placed into 17 redshift bins, each bin having a width
of $\delta z=0.1$. This should be wide enough that observational
uncertainties are uncorrelated from bin to bin.  The parameter analysis
turns out to be not very sensitive to the exact number of supernovae
within each bin, as long as the entire redshift range is reasonably
represented \cite{fhlt}.  The SNAP baseline distribution is tabulated in
\cite{klmm}; generically one also includes data from 300 expected supernovae of
the currently running Nearby Supernova Factory \cite{snf}, as a single
point in the lowest bin.

The magnitude error $\sigma(m_k)$ comprises two contributions: a
statistical component from observational and intrinsic supernova
magnitude dispersion and a systematic component from observational
uncertainties.  These are added in quadrature: \BE \label{master-sigma}
\sigma(m_k)=\sqrt{\frac{\sigma_0^2}{n_k}+\left(m_{sys}\frac{z_k}
{1.7}\right)^2}, \EE where ${k=1,\dots,17}$ labels the 17 bins,
$\sigma_0=0.15$ mag is the statistical calibrated uncertainty of an
individual supernova, $n_k$ is the number of supernovae in bin $k$, and
the redshift $z_k$ is taken at the bin center, i.e.\ at $z_k=0.05$ for
$k=1$.  The second, systematic term  uses the error model discussed in \cite{klmm}, with a linear rise with redshift.  This serves as an approximation of observational uncertainties. SNAP instrumentation and observing strategy is
specifically designed to limit systematic uncertainties to below
$m_{sys}=0.02$ mag out to the maximum redshift $z=1.7$.

The data is related to the theoretical parameters by 
\BA
m(z)&=&5\log_{10}\Biggl[(1+z)\int_0^z dz'
\Bigl[(1-\Omega_D)
(1+z')^3 \nonumber \\
&+&\Omega_De^{3{\int_0^{\ln(1+z')}d(\ln(1+z''))
[1+w(z'')]}}\Bigr]^{-1/2}\Biggr]+{\cal M} \ . \nonumber \EA 
While the use of only one constant parameter $w$ to model the function $w(z)$, i.e.\ assuming a constant
equation of state, is physically unrevealing and often
a poor approximation, more than two parameters makes it
difficult to obtain meaningful constraints. We will write  equation for $m(z)$ using the fit (\ref{fit}) containing two parameters, $w_0$ and $w_a$: 
\BA m(z)&=&5\log_{10}\Biggl[(1+z)\int_0^z dz' \Bigl[(1-\Omega_D) (1+z')^3
\nonumber\\ &+&\Omega_D(1+z')^{3(1+w_0+w_a)}
e^{-3w_a\frac{z'}{1+z'}}\Bigr]^{-1/2}\Biggr]+{\cal M} \ .
\nonumber
\EA

Note that the offset parameter ${\cal
M}=M+25-5\log_{10}(H_0/100 {\rm km/s/Mpc})$ (where $M$ is the supernova
absolute magnitude) just adds linearly, so the sensitivity
derivative ${\partial m_k\over \partial {\cal M}}=1$ for all bins.
Technically, other astrophysical terms enter into $m(z)$
such as a change in supernova magnitude due to dimming by
intervening dust, but we assume that we are dealing with
fully calibrated data, with only the cosmological
dependences remaining.

In Fig.\ \ref{star72v1}  we evaluate 
the sensitivity derivatives at the fiducial
parameter values: $\Omega_D=0.72$, $w_0=-1$, $w_a=0$,
corresponding to the cosmological constant model for the
dark energy.  For the case of only supernova data we also
add a gaussian prior of $\sigma_{\Omega_D}=0.03$,
corresponding to a reasonable level of knowledge on this
parameter by the time SNAP data is available.  Now we have
all the elements needed to
generate the Fisher matrix.  Next we invert it to obtain
the covariance matrix and the parameter estimation
uncertainties (the diagonal elements).  If we want to
plot the confidence contours in a two dimensional section
of the parameter phase space, e.g.\ the $w_0-w_a$ plane,
then we marginalize over the other parameters and reinvert
the covariance matrix to get a reduced, $2\times 2$ Fisher
matrix.  This gives the major and minor axes of the
elliptical contour (it is always an ellipse within the
Fisher approximation) and the orientation, i.e.\ the
direction of the degeneracy between the parameters.  This
is basically an eigenvector where a particular combination
of the parameters is best determined while an orthogonal
combination is poorly constrained.

A plotting program
takes the eigenvector information and draws the ellipse,
with the scale determined by what level of probability
one wants to enclose within the contour.  For 68\% of
the probability enclosed (called $1\sigma$ joint probability),
each axis of the figure is 1.52 ($\sqrt{2.30}$) times larger
than the individual $1\sigma$ probabilities (called
$1\sigma$ projected probability, or enclosing 39\% joint
probability) from the diagonal elements of the covariance
matrix.  For 95\% confidence level, the scaling is 2.45
($\sqrt{5.99}$; formally the numbers 2.30 and 5.99 are the
increments in the $\chi^2$ statistic from the fiducial
model to a model on the 68\%, resp.\ 95\% contours).  As
found here, without a biasing systematic magnitude error,
the best fit model will always be the fiducial, input model.

\subsection*{Planck}
The cosmic microwave background data carries much
information on the cosmological model parameters.  Here
we consider a simple subset of the measurements, involving
only the distance to the photon last scattering surface,
which is exquisitely determined by the location of the
acoustic peaks in the CMB power spectrum. The Planck
Surveyor mission will also precisely determine the
combination $\Omega_M h^2$, so we employ the reduced
distance
$$
\tilde d = \int\limits_0^{z_{{}_{LSS}}} dz \,  f(z)^{-1/2}, 
$$
where the upper boundary of the integral is taken at
the last scattering surface (LSS), $z_{{}_{LSS}}=1089$, and where
$$f(z)=\left[
(1+z)^3+\frac{\Omega_D}{1-\Omega_D}(1+z)^{3(1+w_0+
w_a)}e^{-3w_a\frac{z}{1+z}}\right]$$

Planck has the ability to determine $\tilde d$ up to
0.7\%, i.e.\ $\sigma_{\tilde d}=0.007\cdot\tilde d$; this is translated into a Fisher matrix
\BE
F^{\mathrm{Planck}}_{ij}=\frac{1}{\sigma_{\tilde d}^2}
\frac{\partial \tilde d}{\partial p_i}\frac{\partial
\tilde d}{\partial p_j},
\EE
where $\mathbf{p}=\{\Omega_D, w_0, w_a\}$. Since we want to add
this Fisher matrix to the Fisher matrix from SNAP, we
will include also a fourth column and row of zeros
(for the supernovae parameter ${\cal M}$ that does not
enter the CMB data).

The final Fisher matrix for SNAP and Planck combined
then is
\BE
F^{\mathrm {SNAP+Planck}}= F^{\mathrm {SNAP}}+ F^
{\mathrm {Planck}}.
\EE
Notice that it does not matter that the Fisher
matrices are based on different data quantities from
different experiments ($m(z)$ and $\tilde d$). The
information still adds.  When including the Planck
data, we can eliminate the step of adding a prior on
$\Omega_D$, since the different cosmological quantities measure 
sufficiently different combinations of $\Omega_D$ with the other parameters. 
This breaks degeneracies well enough to determine $\Omega_D$
with superior precision (roughly equivalent to a prior
of $\sigma_{\Omega_D}=0.01$ \cite{fhlt}).

\subsection*{Weak lensing}
The procedure for other data sets, such as from
measurements of the gravitational distortion of images of background 
galaxies by foreground mass concentrations, known as WL, is implemented
similarly.  Here we used only estimates of the future precision of measurements 
of the linear part of the lensing
shear power spectrum (see \S3.3 of \cite{lj03} for details).
An even stronger data set employing the full power spectrum, and possibly other 
lensing methods, should be available from the SNAP wide field survey and
other experiments.


\begin{thebibliography}{4}
\bibitem{numbers} The present density of the universe is given by $\rho_0 =1.8791*h^2 \times 10^{-29}$ g/cm$^{3} \approx 0.95\times 10^{-29}$ g/cm$^{3}$ for $h = 0.71$. Here $H_0 = 100\ h$\ km\ sec$^{-1}$\ Mps$^{-1}$ is the present value of the cosmological constant. In terms of the Planck density $M_p^4 =(8\pi G)^{-2}$, one has $\rho_0 =1.23 \times 10^{-120} M_p^4$, for $h = 0.71$.
\end{thebibliography}
\end{document}